\documentclass[twocolumn,showpacs,preprintnumbers,amsmath,amssymb,prb]{revtex4}

\usepackage{graphicx}
\usepackage{color}

\begin{document}

\title{Thermal properties of graphene under tensile stress}
\author{Carlos P. Herrero}
\author{Rafael Ram\'irez}
\affiliation{Instituto de Ciencia de Materiales de Madrid,
         Consejo Superior de Investigaciones Cient\'ificas (CSIC),
         Campus de Cantoblanco, 28049 Madrid, Spain }
\date{\today}

\begin{abstract}
Thermal properties of graphene display peculiar characteristics
associated to the two-dimensional nature of this crystalline membrane. 
These properties can be changed and tuned in the presence of
applied stresses, both tensile and compressive.
Here we study graphene monolayers under tensile stress 
by using path-integral molecular dynamics (PIMD) simulations, which 
allows one to take into account quantization of vibrational modes
and analyze the effect of anharmonicity on physical observables.
The influence of the elastic energy due to strain in the crystalline
membrane is studied for increasing tensile stress and for
rising temperature (thermal expansion).
We analyze the internal energy, enthalpy, and specific heat of graphene,
and compare the results obtained from PIMD simulations with those 
given by a harmonic approximation for the vibrational modes. 
This approximation turns out to be precise at low temperatures, and
deteriorates as temperature and pressure are increased.
At low temperature the specific heat changes as $c_p \sim T$ for
stress-free graphene, and evolves to a dependence $c_p \sim T^2$ as the
tensile stress is increased.
Structural and thermodynamic properties display nonnegligible quantum
effects, even at temperatures higher than 300~K.
Moreover, differences in the behavior of the in-plane and real areas 
of graphene are discussed, along with their associated properties. 
These differences show up clearly in the corresponding 
compressibility and thermal expansion coefficient.
\end{abstract}



\maketitle

\section{Introduction}

Two-dimensional materials, and graphene in particular, have attracted
in last years a great deal of attention in the scientific community,
due to their peculiar electronic,
elastic, and thermal properties.\cite{ge07,ro17,ca09b}
Thus, graphene presents high values of the thermal
conductivity,\cite{ba11,gh08b} and large in-plane elastic
constants.\cite{le08} Moreover, its mechanical properties are interesting for
possible applications, as cooling of electronic devices.\cite{pr10,se10}

The ideal structure of pure defect-free graphene is a planar honeycomb 
lattice, but departures from this flat configuration may appreciably alter 
its atomic-scale and macroscopic properties.\cite{me07}
Several reasons can cause bending of a graphene sheet, such as the presence 
of defects and external stresses.\cite{fa07}
In addition, thermal fluctuations at finite temperatures give rise to
out-of-plane displacements of the C atoms, and for $T \to 0$, 
zero-point motion yields also a departure of strict planarity 
of the graphene sheet.\cite{he16}

The influence of strain in several characteristics of two-dimensional (2D)
materials, such as graphene and metallic dichalcogenides, has been 
emphasized in recent years. This includes electronic transport,
optical properties, and the formation of moir\'e patterns.\cite{wo14,am16}  
From a structural viewpoint, external stresses may cause significant 
changes in crystalline membranes, which can crumple in the presence of 
a compressive stress, as is well known for lipid membranes\cite{ev90,sa94} 
and polymer films.\cite{ce03,wo06}
For graphene, crumpling was observed in supported as well as freestanding 
samples, and has been explained as due to both out-of-plane phonons and
static wrinkling.\cite{ki11,ni15} 
Recent molecular dynamics simulations indicate that the maximum compressive 
stress that a freestanding graphene sheet can sustain without crumpling 
decreases as the system size grows, and was estimated to be about 0.1 N/m 
at room temperature in the thermodynamic limit.\cite{ra17}

A tensile stress in the graphene plane does not break
the planarity of the sheet, but gives rise to significant variations 
in the elastic properties of the material.\cite{lo17}
For example, the in-plane Young modulus increases by a factor of three 
for a tensile stress of 1 N/m.\cite{ra17}
In this respect, the real area per atom, $A$, can be crucial to
understand the elastic properties, which have been in the past 
usually referred to its projection, $A_p$, onto the mean plane of 
the membrane ($A_p \leq A$). This question has been discussed along
the years for biological membranes,\cite{he84,sa94,fo08,ta13,ch15} 
and has been recently examined for crystalline membranes such 
as graphene.\cite{po11b,ni15,ni17,ra16,he16}
In particular, the high-quality data obtained by 
Nicholl {\em et al.}\cite{ni15,ni17} clearly indicate that some 
experimental techniques can measure properties related to the real
area $A$, whereas other techniques may be used to quantify variables
connected to the projected area $A_p$.
The difference $A - A_p$ has been called {\em hidden} area for graphene
in Ref.~\onlinecite{ni17}, as well as {\em excess} area for biological 
membranes.\cite{he84,fo08}
A precise knowledge of the behavior of both areas is important to
clarify the temperature and stress dependence of structural and
thermodynamic properties of graphene.
Thus, it is possible to define the elastic properties in
relation to the area $A$ or to $A_p$, which may behave very differently.
In fact, it is known that referring to the in-plane area $A_p$, 
one finds a negative thermal expansion coefficient, but for the real 
area $A$ the thermal expansion is positive.\cite{po11b,he16}

A deep comprehension of the properties of 
2D systems has been for many years a persistent goal in
statistical physics.\cite{sa94,ne04,ta13}
This has been in part due to the complexity of the considered
systems, such as biological membranes and soft condensed 
matter.\cite{ta13,ch15}
In this respect, graphene can be dealt with as a model system where
descriptions at an atomic level can be connected with physical
properties of the material.
Thus, thermal properties of graphene have been investigated
in recent years,\cite{am14,po12b,wa16,fo13,ma16}
and in particular its thermal expansion and heat conduction
were studied by various theoretical and experimental
techniques.\cite{ba11,po12b,ma12,al13,si14,bu16}

Several theoretical works carried out to study thermodynamic
properties of graphene (e.g., specific heat, thermal expansion, ...),
were based on density-functional-theory (DFT) calculations combined with
a quantum quasi-harmonic approximation (QHA) for 
the vibrational modes.\cite{mo05,se14,ma17}
This is expected to yield reliable results at low temperature, but
may be questioned at relatively high temperature, especially for the
thermal expansion, due to an important anharmonic coupling between
in-plane and out-of-plane modes, not included in the QHA.
Moreover, classical Monte Carlo and molecular dynamics simulations based
on {\em ab-initio},\cite{sh08,po11b,an12b,ch14}
tight-binding,\cite{he09a,ca09c,ak12,le13}
and empirical potentials\cite{fa07,ra16,ma14,lo16}
can give reliable results at relatively high temperature, 
but fail to describe thermodynamic properties at $T < \Theta_D$, with
$\Theta_D \gtrsim$ 1000 K the Debye temperature of the material.
This means, in particular, that room-temperature results obtained for 
some properties of graphene from classical atomistic simulations may be 
clearly unrealistic.
These shortcuts may be overcome by using simulation methods which
explicitly include nuclear quantum effects, in particular those based 
on Feynman path integrals.\cite{gi88,ce95,br15,he16}

Here we use the path-integral molecular dynamics (PIMD) method 
to study thermal properties of graphene under tensile stress at temperatures
between 12 and 2000~K.
The thermal behavior of the graphene surface is studied, considering
the difference between in-plane and real areas.
The in-plane thermal expansion coefficient turns out to be negative
at low temperatures, with a crossing to positive values at a temperature
which decreases fast as tensile stress is raised.
Particular emphasis is laid on the temperature dependence of the specific
heat at low $T$, for which results of the simulations are compared with
predictions based on harmonic vibrations of the crystalline membrane.
This approximation happens to be noticeably accurate at low temperatures, 
once the frequencies of out-of-plane ZA modes are properly renormalized
for changing applied stress.

 The paper is organized as follows. In Sec.\,II we describe the
computational method employed in the simulations.
Results for the internal energy and enthalpy of graphene are given 
in Sec.~III.  The thermal expansion is discussed in Sec.~IV.
In Sec.~V we present results for the specific heat, 
whereas in Sec.\,VI the compressibility of graphene is discussed.  
In Sec.\,VII we summarize the main results.

\vspace{1cm}

\section{Computational Method}

\subsection{Path-integral molecular dynamics}

We use PIMD simulations to study equilibrium properties of 
graphene monolayers as a function of temperature and pressure.
The PIMD procedure is based on the Feynman path-integral formulation of
statistical mechanics,\cite{fe72} an adequate nonperturbative approach 
to study finite-temperature properties of many-body quantum systems.
In the applications of this method to numerical simulations,
each quantum particle is represented by a set of $N_{\rm Tr}$ 
(Trotter number) replicas (or beads), that behave as classical-like
particles forming a ring polymer.\cite{gi88,ce95}
Thus, one deals with a {\em classical isomorph} whose dynamics is 
artificial, since it does not reflect the real quantum dynamics of 
the actual particles, but is useful for effectively sampling the
many-body configuration space, yielding precise results for
time-independent equilibrium properties of the quantum system.
Details on this type of simulation techniques are given
in Refs.~\onlinecite{gi88,ce95,he14,ca17}.

The interatomic interactions between carbon atoms are described here
by using the LCBOPII effective potential,
a long-range bond order potential, which has been mainly employed
to carry out classical simulations of carbon-based systems.\cite{lo05} 
In particular, it was used to study the phase diagram of carbon,
including graphite, diamond, and the liquid, and showed its accuracy 
in predicting rather precisely the graphite-diamond 
transition line.\cite{gh05b}
In recent years, this interatomic potential has been also found
to describe well several properties of graphene,\cite{fa07,lo16} 
and its Young's modulus in particular.\cite{za09,po12,ra17} 
This interatomic potential was lately employed to carry out PIMD
simulations, which allowed to quantify quantum effects
in graphene monolayers by comparing with results of classical
simulations,\cite{he16} as well as to study thermodynamic properties
of this 2D material.\cite{he18}
Here, in line with earlier simulations,\cite{ra16,he16,ra17} 
the original parameterization of the LCBOPII potential
has been slightly changed to rise the zero-temperature 
bending constant $\kappa$ from 1.1 eV to 1.49 eV, a value close
to experimental data.\cite{la14}

The calculations presented here have been performed in the 
isothermal-isobaric ensemble, where one fixes the number of carbon 
atoms ($N$), the applied stress ($P$), and the temperature ($T$).
The stress $P$ in the reference $xy$ plane of graphene, 
with units of force per unit length, coincides with the
so-called mechanical or frame tension.\cite{ra17,fo08,sh16}
$P$ is obtained in the simulations from the stress tensor 
${\bf \tau}$, whose components are given by expressions
such as\cite{tu98,he18} 
\begin{widetext}
\begin{equation}
 \tau_{xy} = \left \langle \frac{1}{N A_p}
   \left(  \sum_{i=1}^{N}  \sum_{j=1}^{N_{\rm Tr}}
     \left( m_j v_{ij,x} v_{ij,y} - 2 k_j u_{ij,x} u_{ij,y} \right) -  
     \frac{1}{N_{\rm Tr}} \sum_{j=1}^{N_{\rm Tr}}
     \frac {\partial U({\bf r}_{1j}, \dots, {\bf r}_{Nj})}
           {\partial \epsilon_{xy}} \right) \right \rangle  \; ,
\label{tauxy}
\end{equation}
\end{widetext}
where the brackets $\langle \cdots \rangle$ indicate an ensemble average
and ${\bf u}_{ij}$ are staging coordinates,\cite{tu93} with 
$i = 1, ..., N$ and $j = 1, ..., N_{\rm Tr}$.
In Eq.~(\ref{tauxy}), $m_j$ is the dynamic mass associated to 
${\bf u}_{ij}$ and $v_{ij,x}$, $v_{ij,y}$ are components of its
corresponding velocity. 
The constant $k_j$ is given by
$k_j = m_j N_{\rm Tr} / 2 \beta^2 \hbar^2$ for $j > 1$ and $k_1 = 0$.
Here $\beta = (k_B T)^{-1}$, $U$ is the instantaneous potential energy,
and $\epsilon_{xy}$ is an element of the 2D strain tensor.
The stress $P$, conjugate to the in-plane area $A_p$, is given by
the trace of the stress tensor:
\begin{equation}
  P = \frac{1}{2} \left( \tau_{xx} + \tau_{yy} \right)  \, .
\end{equation}

Effective algorithms have been employed for the PIMD simulations, 
as those described in the literature.\cite{tu92,tu98,ma99} 
In particular, a constant temperature $T$ was accomplished by 
coupling chains of four Nos\'e-Hoover thermostats,
and an additional chain of four barostats was coupled to 
the area of the graphene simulation box to yield the required 
stress $P$.\cite{tu98,he14}
The equations of motion were integrated by employing the reversible 
reference system propagator algorithm (RESPA), allowing to define
different time steps for the integration of the fast and slow
degrees of freedom.\cite{ma96}
The kinetic energy $K$ has been calculated by using the virial
estimator, which displays a statistical uncertainty significantly 
smaller than the potential energy, 
$V = \langle U \rangle$.\cite{tu98}
The time step $\Delta t$ associated to the interatomic forces has been
taken as 0.5 fs, which was found to be suitable for the carbon atomic 
mass and the temperature range considered here.
More details on this kind of PIMD simulations can be found
elsewhere.\cite{tu98,he06,he11}

Rectangular simulation cells with $N$ = 960 atoms have been considered, 
with similar side lengths in the $x$ and $y$ directions 
($L_x \approx L_y \approx 50$~\AA),
and periodic boundary conditions were assumed.
Sampling of the configuration space was performed in the temperature 
range between 12~K and 2000~K.
A temperature-dependent Trotter number has been defined as
$N_{\rm Tr} = 6000 {\rm K} / T$, which yields a roughly constant 
precision for the PIMD results at different 
temperatures.\cite{he06,he11,ra12}
Given a temperature $T$ and a stress $P$, a typical simulation run 
included $3 \times 10^5$ PIMD steps for system equilibration and
$6 \times 10^6$ steps for calculation of average properties.

\subsection{In-plane vs real area}

As explained above, in the isothermal-isobaric ensemble used here 
we fix the applied stress $P$ in the $xy$ plane, allowing 
fluctuations in the in-plane area of the simulation
cell for which periodic boundary conditions are applied.
Carbon atoms are free to move in the $z$ coordinate (out-of-plane 
direction), and in general any measure of the {\em real} surface
of a graphene sheet at $T > 0$ should give a value larger than
the in-plane area.
In this respect, it has been discussed for biological membranes that
their properties should be described using the notion 
of a real surface rather than a {\em projected} (in-plane)
surface.\cite{wa09,ch15} 
A similar question has been also recently posed for crystalline membranes
such as graphene.\cite{po11b,he16,ni17,ra17}
This may be relevant for addressing the calculation of thermodynamic 
variables, as the in-plane area $A_p$ is the variable conjugate to the 
stress $P$ used in our simulations. The real area (also called true, 
actual, or effective area in the literature\cite{fo08,wa09,ch15})
is conjugate to the usually-called surface tension.\cite{sa94}

The in-plane area has been most used to present the results of atomistic 
simulations of graphene layers.\cite{ga14,br15,za09,lo16,ch15}
For biological membranes, however, it has been shown that values of 
the compressibility may significantly differ when they are related 
to $A$ or to $A_p$, and
something similar has been recently found for the elastic properties of
graphene from classical molecular dynamics simulations.\cite{ra17}

Here we calculate a {\em real} area $A$ in three-dimensional (3D) space 
by a triangulation based on the actual positions of the C atoms along 
a simulation run (in fact we use the beads associated to the atomic nuclei).
Specifically, $A$ is obtained from a sum of areas corresponding
to the structural hexagons. 
Each hexagon contributes as a sum of six triangles, each one formed 
by the positions of two adjacent C atoms and the barycenter of 
the hexagon.\cite{ra17}

The instantaneous area per atom for {\em imaginary time} 
(bead) $j$ is given by 
\begin{equation}
    A^j =  \frac1N \sum_{k=1}^{2N} \sum_{n=1}^6 T_{kn}^j
\label{aj}
\end{equation}
where $T_{kn}^j$ is the area of triangle $n$ in hexagon $k$,
and the sum in $k$ is extended to the $2N$ hexagons in a cell 
containing $N$ carbon atoms. Here, the triangles are defined
with the coordinates ${\bf r}_{ij}$ corresponding to bead $j$
of atom $i$.
The area $A$ is then calculated as
\begin{equation}
  A =  \left \langle \frac{1}{N_{\rm Tr}} \sum_{j=1}^{N_{\rm Tr}} A^j
        \right \rangle 
\label{aa}
\end{equation}

It is clear that $A$ coincides with $A_p$ for strictly planar 
graphene, and in general one has $A \geq A_p$.
Both areas display temperature dependencies qualitatively
different. In fact, $A$ does not present a negative thermal expansion,
as happens for the in-plane area in a large temperature 
range\cite{za09,he16} (see below). 
The difference $A - A_p$ increases with temperature,
as bending of the graphene sheet increases, but even for $T \to 0$,
$A$ and $A_p$ are not exactly equal, due to zero-point motion of
the C atoms in the transverse $z$ direction.

\section{Energy and enthalpy}

\subsection{Internal energy}

The internal energy $E$ is obtained from the results of our PIMD
simulations as a sum of the kinetic, $K$, and potential energy, $V$, 
at a given temperature.
The kinetic energy has been calculated by employing the virial
estimator,\cite{tu98} which displays a statistical uncertainty
smaller than the potential energy.  We have included in $E$
the center-of-mass translational energy, a classical magnitude 
amounting to $E_{\rm CM} = 3 k_B T / 2$ at temperature $T$. 
This quantity is irrelevant for the energy per atom in large systems, 
but we have included it to minimize finite size effects.\cite{he16} 

\begin{figure}
\vspace{-0.7cm}
\includegraphics[width=8.0cm]{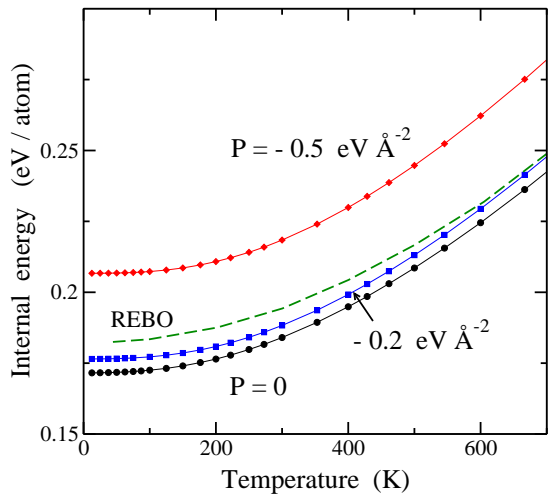}
\vspace{-0.5cm}
\caption{Internal energy per atom $E - E_0$ as a function of temperature
for $P = 0$ (circles), $-0.2$ (squares), and $-0.5$ eV \AA$^{-2}$
(diamonds).  Symbols represent results of PIMD simulations.
Error bars are smaller than the symbol size.
Solid lines are guides to the eye.
The zero-point energy amounts to 172 meV/atom for
$P = 0$ and 207 meV/atom for $P = -0.5$ eV \AA$^{-2}$.
The dashed line corresponds to the results obtained by Brito
{\em et al.}\cite{br15} using the REBO potential.
Note that 1 eV \AA$^{-2}$ = 16 N m$^{-1}$ in SI units.
}
\label{f1}
\end{figure}

We express the internal energy as $E = E_0 + V + K$,
taking as energy reference the value $E_0$ corresponding to 
the equilibrium configuration of a planar graphene surface in a 
classical approach at $T = 0$ (minimum-energy configuration of 
the considered LCBOPII potential, without quantum atomic delocalization).
This corresponds to an interatomic distance $d_{\rm C-C} = 1.4199$ \AA, 
i.e., an area $A_0 = 2.6189$ \AA$^2$ per atom.
In a quantum approach, out-of-plane atomic fluctuations associated 
to zero-point motion appear even for $T \to 0$,
and the graphene layer is not strictly planar. 
Moreover, anharmonicity of in-plane vibrations gives rise to a zero-point 
lattice expansion, yielding a distance $d_{\rm C-C} = 1.4287$ \AA,
i.e., around 1\% larger than the classical distance at $T = 0$.\cite{he16}

In Fig.~1 we show the temperature dependence of the internal energy
per atom, $E - E_0$, as derived from our PIMD simulations in the 
isothermal-isobaric ensemble for $P$ = 0 (circles), $-0.2$ (squares), 
and $-0.5$ eV \AA$^{-2}$ (diamonds). 
For $P = 0$ it was shown earlier that the size effect on the internal 
energy per atom is negligible for $N$ = 960 (less than the symbol
size in Fig.~1), as compared with the largest cells 
considered in Ref.~\onlinecite{he16}.
The zero-point energy, $E_{\rm ZP}$, is found to be 172 meV/atom
for $P$ = 0, and slightly higher for $-0.2$ eV \AA$^{-2}$, 
but it appreciably increases for $P = -0.5$ eV \AA$^{-2}$ to a value 
of 207 meV/atom.
We will show later that this rise in internal energy is basically due
to the elastic energy associated to an increase in the area $A$.
For comparison with our results, we also present in Fig.~1 the energy
$E - E_0$ obtained by Brito {\em et al.}\cite{br15} from path-integral
Monte Carlo simulations using the REBO potential (dashed line). 
These authors called it vibrational energy, and corresponds 
to our internal energy; our vibrational energy $E_{\rm vib}$ is
defined below (see Sec.~III.C).
The REBO potential yields a zero-point energy of
0.181 eV/atom, i.e., a 5\% higher than that found here with the
LCBOPII potential.

\begin{figure}
\vspace{-0.7cm}
\includegraphics[width=8.0cm]{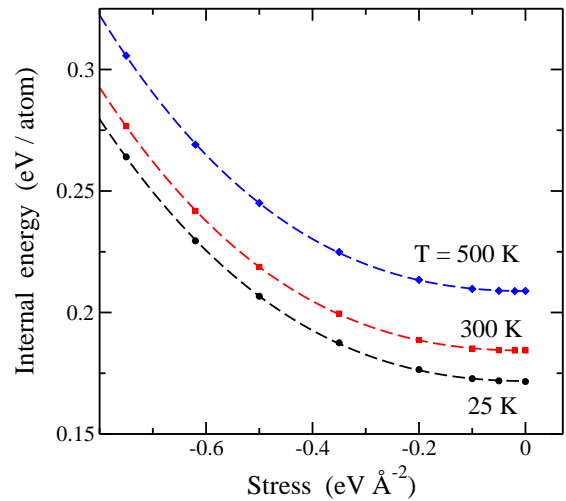}
\vspace{-0.5cm}
\caption{Stress dependence of the internal energy of graphene for
three temperatures: $T$ = 25 (circles), 300 (squares), and 500 K
(diamonds).
Symbols indicate simulation results and lines are fits to the
expression $E - E_0 = c_0 + c_2  P^2 + c_3  P^3$.
}
\label{f2}
\end{figure}

The rise in internal energy with an applied tensile stress is 
presented in Fig.~2 for three temperatures: $T$ = 25 K (circles),
300 K (squares), and 500 K (diamonds).
For zero stress we find at 25 K an internal energy very close to that
of the ground state $E_{\rm ZP}$. 
In the stress range displayed in Fig.~2 the internal energy
can be fitted to an expression $E - E_0 = c_0 + c_2 P^2 + c_3 P^3$,
with a coefficient of the quadratic term 
$c_2 \approx 0.090$ \AA$^4$ eV$^{-1}$, nearly independent 
of the temperature. This is the leading term for the variation of
internal energy with the applied pressure, which is nearly parabolic
for stresses between 0 and $-0.1$ eV \AA$^{-2}$.

The change in internal energy with temperature and pressure can be
described from variations in the elastic and vibrational contributions
to $E$. This is analyzed in the following sections.

\subsection{Elastic energy}

An important part of the internal energy corresponds to the
elastic energy due to changes in the area of graphene.
It is directly related to the actual interatomic distance, i.e.,
to the real area $A$, rather than to the in-plane area $A_p$
(or in-plane strain $\epsilon$).
Then, the internal energy $E(T)$ at temperature $T$ can be 
written as\cite{he16}
\begin{equation}
    E(T) =  E_0 + E_{\rm el}(A) + E_{\rm vib}(A,T)   \, ,
\label{et}
\end{equation}
where $E_{\rm el}(A)$ is the elastic energy for an
area $A$, and $E_{\rm vib}(A,T)$ is the vibrational energy of the system.
The area $A$ is a function of the stress $P$ and temperature $T$, 
but this is not explicitly indicated in Eq.~(\ref{et}) for simplicity
of the notation.
One expects non-zero values of the elastic energy, even in the absence 
of an externally applied stress, due to thermal expansion at finite
temperatures, as well as for zero-point expansion at $T = 0$.
$E_{\rm vib}(A,T)$ is given by contributions of phonons in graphene,
both in-plane and out-of-plane vibrational modes.

\begin{figure}
\vspace{-0.7cm}
\includegraphics[width=8.0cm]{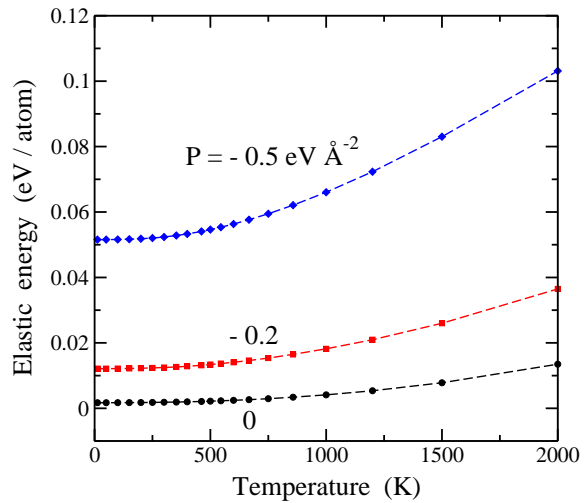}
\vspace{-0.5cm}
\caption{Temperature dependence of the elastic energy $E_{\rm el}$
of graphene, as derived from the real area $A$ obtained in PIMD
simulations for $P = 0$ (circles), $-0.2$ (squares), and
$-0.5$ eV \AA$^{-2}$ (diamonds).
Symbols represent simulation results and error bars are less than
the symbol size.
Dashed lines are guides to the eye.
}
\label{f3}
\end{figure}

We define the elastic energy $E_{\rm el}$ corresponding to an area $A$ 
as the increase in energy of a strictly planar graphene layer with 
respect to the minimum energy $E_0$ 
(for an area $A_0$ = 2.61888 \AA$^2$/atom).
By definition $E_{\rm el}(A_0) = 0$, and for small changes of $A$
we found that it follows a dependence
$E_{\rm el}(A) \approx K (A - A_0)^2$, with $K$ = 2.41 eV \AA$^{-2}$.
This dependence for $E_{\rm el}(A)$ yields a 2D bulk modulus
$B_0 = A_0 ( \partial^2 E_{\rm el} / \partial A^2 )$ = $2 K A_0$,
i.e., $B_0$ = 12.6 eV \AA$^{-2}$, in agreement with the result found
earlier in classical calculations at $T = 0$.\cite{ra17} 

Our PIMD simulations directly yield $E(T)$, which can be split
into an elastic and a vibrational part, as in Eq.~(\ref{et}).
At room temperature ($T \sim$ 300~K) and for small stresses $P$
($A$ close to $A_0$), the elastic energy is much smaller than
the vibrational energy $E_{\rm vib}$, but this can be different for
low $T$ and/or large applied stresses (see below).

In Fig.~3 we display the temperature dependence of the elastic energy,
as derived from our PIMD simulations for $P = 0$, $-0.2$, and
$-0.5$ eV \AA$^{-2}$.
For a given external stress, $E_{\rm el}$ increases with $T$, as
a consequence of the thermal expansion of the real area $A$,
which turns out to be positive for all temperatures $T > 0$
($\alpha > 0$, see below). Note that, in contrast, the in-plane
area $A_p$ displays a thermal contraction ($\alpha_p < 0$) in a
wide temperature range, so that it is not a suitable candidate for
a reliable definition of the elastic energy.

For $P$ = 0 we find a positive elastic energy in the 
zero-temperature limit, $E_{\rm el}$ = 1.7 meV/atom, due to zero-point 
lattice expansion, which causes that $A > A_0$.
This low-$T$ limit appreciably increases for $P < 0$ due to
the stress-induced increase in area $A$, which yields values of
12 and 52 meV/atom for the elastic energy at $P = -0.2$ and 
$-0.5$ eV~\AA$^{-2}$, respectively.
For  finite temperatures, one observes in Fig.~3 that 
$E_{\rm el}(T)$ increases faster with temperature for larger tensile 
stress.  This is due to an increase in the graphene compressibility 
(reduction of the elastic constants) for rising tensile stress, which
in turn causes an increase in the thermal expansion coefficient $\alpha$
(see Sec.~IV).

\begin{figure}
\vspace{-0.7cm}
\includegraphics[width=8.0cm]{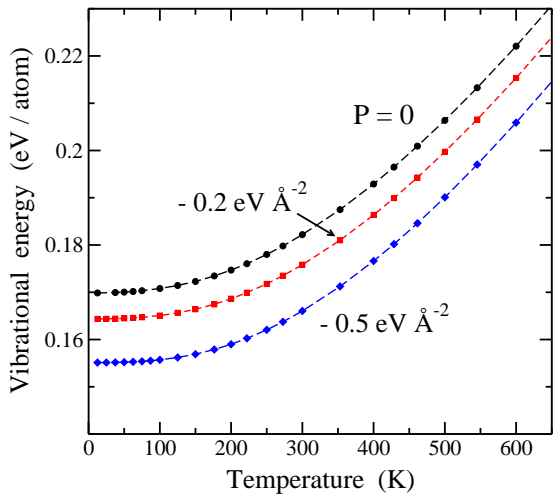}
\vspace{-0.5cm}
\caption{Temperature dependence of the vibrational energy per atom,
$E_{\rm vib}$, for $P = 0$ (circles), $-0.2$ (squares), and
$-0.5$ eV \AA$^{-2}$ (diamonds).
Error bars are less than the symbol size.
Dashed lines are guides to the eye.
}
\label{f4}
\end{figure}

\subsection{Vibrational energy}

After calculating the elastic energy for an area $A$ resulting from
PIMD simulations at given $T$ and $P$,
we obtain the vibrational energy $E_{\rm vib}(A,T)$ by subtracting
the elastic energy from the internal energy:
$E_{\rm vib} = E(T) - E_0 - E_{\rm el}(A)$ (see Eq.~(\ref{et})).
In Fig.~4 we present the temperature dependence of the vibrational 
energy of graphene for $P = 0$ (circles), $-0.2$ (squares), and
$-0.5$ eV \AA$^{-2}$ (diamonds), as derived from our simulations.
In contrast to Figs.~1 and 3, where one observes that $E$ and 
$E_{\rm el}$ increase with the applied tensile stress $P$, in
Fig.~4 one sees that the vibrational energy is lower for
higher tensile stress.
This is mainly due to a decrease in the vibrational frequency
of in-plane modes for increasing stress (increasing area),
corresponding to positive Gr\"unesien parameters.\cite{as76,mo05}

The zero-point vibrational energy is found to decrease from 
170 meV/atom for $P = 0$ to 164 and 155 meV/atom for $P = -0.2$ and 
$-0.5$ eV \AA$^{-2}$, respectively. 
This decrease, although clearly noticeable, is
smaller than the increase of about 50 meV/atom in the elastic energy
for $P = -0.5$ eV \AA$^{-2}$ (see Fig.~3).
These zero-$T$ energy values per atom correspond in a harmonic 
approximation to $\langle 3 \hbar \omega / 2 \rangle$, 
which is a mean value for the frequencies $\omega$ in the six phonon
bands of graphene.\cite{ka11,wi04}
Our results for $P = 0$ correspond to 
$\langle \omega \rangle = 914$ cm$^{-1}$, to be compared with
833 cm$^{-1}$ for $P = -0.5$ eV \AA$^{-2}$.

The three curves for $E_{\rm vib}(T)$ shown in Fig.~4 for different 
stresses are roughly parallel in the displayed temperature range.
Values of $E_{\rm vib}$ presented in this figure are clearly
larger than those corresponding to a classical model for the
vibrational modes. In this limit, the vibrational energy per
atom is $E_{\rm vib}^{\rm cl} = 3 k_B T / 2$, which means
39 and 78 meV/atom for $T = 300$ and 600 K, respectively. 
In contrast, we find $E_{\rm vib}$ = 166 and 206 meV/atom from
PIMD simulations for $P = -0.5$ eV \AA$^{-2}$ at those temperatures.

We note that in the language of membranes and 2D elastic media
 there appears an energy contribution due to bending of the surface,
that is usually taken into account
through the bending constant $\kappa$, which measures the
rigidity of the membrane. In our present formulation, the bending energy
is included in the vibrational energy associated to the flexural ZA modes,
as can be seen in their contribution to the specific heat (see Sec.~V). 
Anharmonic couplings between in-plane and out-of-plane modes are also
expected to show up, especially at high temperatures.\cite{ad16}

\begin{figure}
\vspace{-0.7cm}
\includegraphics[width= 8.0cm]{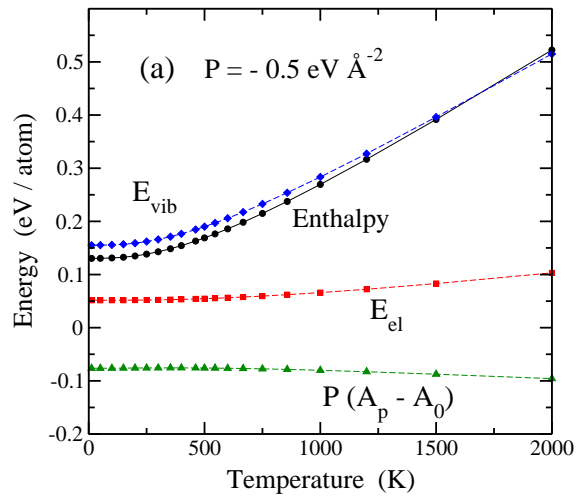}
~\vspace{-2.0cm}
\includegraphics[width= 8.0cm]{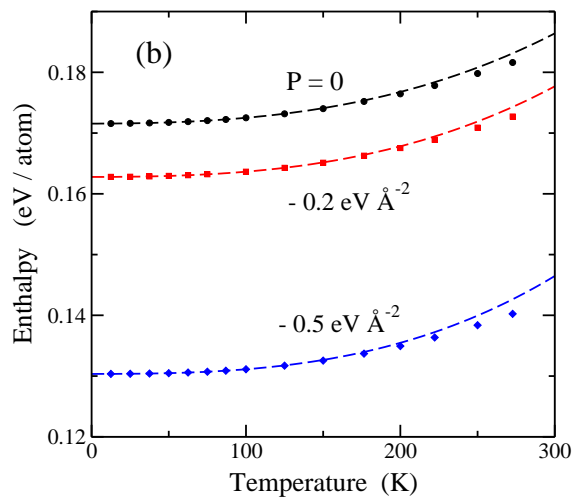}
\vspace{1.5cm}
\caption{Enthalpy per atom, $H - H_0$, as a function of temperature.
(a) Enthalpy for a tensile stress $P = -0.5$ eV \AA$^{-2}$ (circles),
along with its three components, $E_{\rm vib}$, $E_{\rm el}$, and
$P (A_p - A_0)$ (see Eq.~(\ref{hh0})).
Dashed lines are guides to the eye.
(b) Enthalpy in the low-temperature region for $P = 0$ (circles),
$-0.2$ (squares), and $-0.5$ eV \AA$^{-2}$ (diamonds).
Symbols represent results of PIMD simulations.
Error bars are less than the symbol size.
Dashed lines in (b) indicate fits of the simulation results to the
polynomial $H - H_0(P) = H_{\rm ZP} + a_2 T^2 + a_3 T^3$ in the region
from $T = 0$ to 100 K.
}
\label{f5}
\end{figure}

\subsection{Enthalpy}

Since we are working in the isothermal-isobaric ensemble,
it is natural to consider the enthalpy as a relevant thermodynamic
variable, in particular to calculate the specific heat
of graphene at several applied stresses (see Sec.~V).
For a given stress $P$, we define the enthalpy $H$ as
$H =  E + P A_p$. 
As a reference we will consider $H_0(P) = E_0 + P A_0$,
such that $H$ converges in the classical limit
to $H_0(P)$ at $T = 0$.

In Fig.~5(a) we present the enthalpy of graphene, $H - H_0(P)$, 
as a function of temperature for a tensile stress 
$P = -0.5$ eV~\AA$^{-2}$. Results of our simulations are displayed as
circles. Taking into account that the enthalpy can be written as
\begin{equation}
   H - H_0(P) = E_{\rm el} + E_{\rm vib} + P (A_p - A_0)  \; ,
\label{hh0}
\end{equation}
we have plotted in Fig.~5(a) the temperature dependence of the
three contributions in the r.h.s. of Eq.~(\ref{hh0}).
The largest change with temperature appears for the vibrational energy
(diamonds). The elastic energy, $E_{\rm el}$, as well as the term
$P (A_p - A_0)$ are also found to depend on temperature, 
due to the thermal expansion (or contraction) of the in-plane area $A_p$, 
but their change is much smaller than that of 
$E_{\rm vib}$.  As a result, the enthalpy turns out to be smaller than
the vibrational energy at low temperature, because
$E_{\rm el} + P (A_p - A_0) < 0$. 
However, $H - H_0(P)$ becomes larger than $E_{\rm vib}$ at 
$T \gtrsim 1500$~K, due to the larger increase in $E_{\rm el}$ for
rising temperature. 

To better appreciate the low-temperature region,
in Fig.~5(b) we present the temperature dependence of the enthalpy,
as derived from PIMD simulations up to $T = 300$ K.
Symbols indicate results of the simulations for $P$ = 0 (circles),
$-0.2$ (squares), and $-0.5$ eV \AA$^{-2}$ (diamonds)
Dashed lines are fits to the expression
$H - H_0(P) = H_{\rm ZP} + a_2 T^2 + a_3 T^3$ in the temperature 
region from $T = 0$ to 100~K. 
This polynomial form shows good agreement with the temperature 
dependence of the enthalpy obtained from the simulations in
the fitted region. For $T \gtrsim 150$~K, the lines depart 
progressively from the data points. 
Note that a linear term in this expression for the enthalpy is
not allowed for thermodynamic consistency, since the specific heat
$c_p = (\partial H / \partial T)_P$ has to vanish
in the limit $T \to 0$.
The coefficient $a_2$ changes from $6.1 \times 10^{-8}$ eV K$^{-2}$ 
for $P = 0$ to $2.6 \times 10^{-8}$ eV K$^{-2}$ for 
$P = -0.5$ eV \AA$^{-2}$.  
The coefficient $a_3$ varies from 3.5 to 
$5.1 \times 10^{-10}$ eV K$^{-3}$ in the same stress range.
This is important for the low-temperature dependence of the specific
heat discussed in Sec.~V, as the contribution of the quadratic
coefficient $a_2$ decreases for increasing stress, whereas 
$a_3$ is found to increase.
This means that the specific heat $c_p$ in the lowest temperatures 
accessible to our simulation method will be given as a combination
of a linear term $2 a_2 T$ and a quadratic one $3 a_3 T^2$, with the
latter becoming more important for larger tensile stresses.

\section{Thermal expansion}

In the low-temperature limit ($T \to 0$), the real area $A$ and 
in-plane area $A_p$ derived from PIMD simulations converge to 
2.6459 \AA$^2$/atom and 2.6407 \AA$^2$/atom, respectively. 
Comparing these values with the classical minimum $A_0 = 2.6189$ \AA,
we find a zero-point expansion in $A$ and $A_p$ of about 
0.02 \AA$^2$/atom ($\sim$ 1\%), 
due to an increase in the mean C--C bond length (an anharmonic effect). 
There also appears a difference of 0.2\% between real and in-plane 
areas at low temperature, caused by out-of-plane zero-point motion, 
so that the layer is not strictly planar. 
This is a genuine quantum effect, as in classical simulations 
for $T \to 0$ one finds a planar layer in which $A$ and $A_p$ 
coincide.\cite{he16,ra17}
The difference $A - A_p$ increases as temperature is raised, since 
$A_p$ is the projection of $A$ on the $xy$ reference plane, 
and the real graphene surface becomes increasingly bent for rising 
temperature because of larger out-of-plane atomic displacements.

For $P = 0$, the area $A$ displays an almost constant value up to 
$T \approx 200$ K, and increases at higher temperatures. 
However, $A_p$ decreases in the temperature range from $T = 0$ to 
$T \approx 1000$ K, reaches a minimum and then increases 
at higher $T$.\cite{he16}
These results for $A_p$ are qualitatively similar to those found
from classical Monte Carlo and molecular dynamics simulations
of graphene,\cite{za09,ga14} but in PIMD simulations the contraction 
of $A_p$ with respect to the zero-temperature value is significantly 
larger than for classical calculations.

\begin{figure}
\vspace{-0.7cm}
\includegraphics[width= 8.0cm]{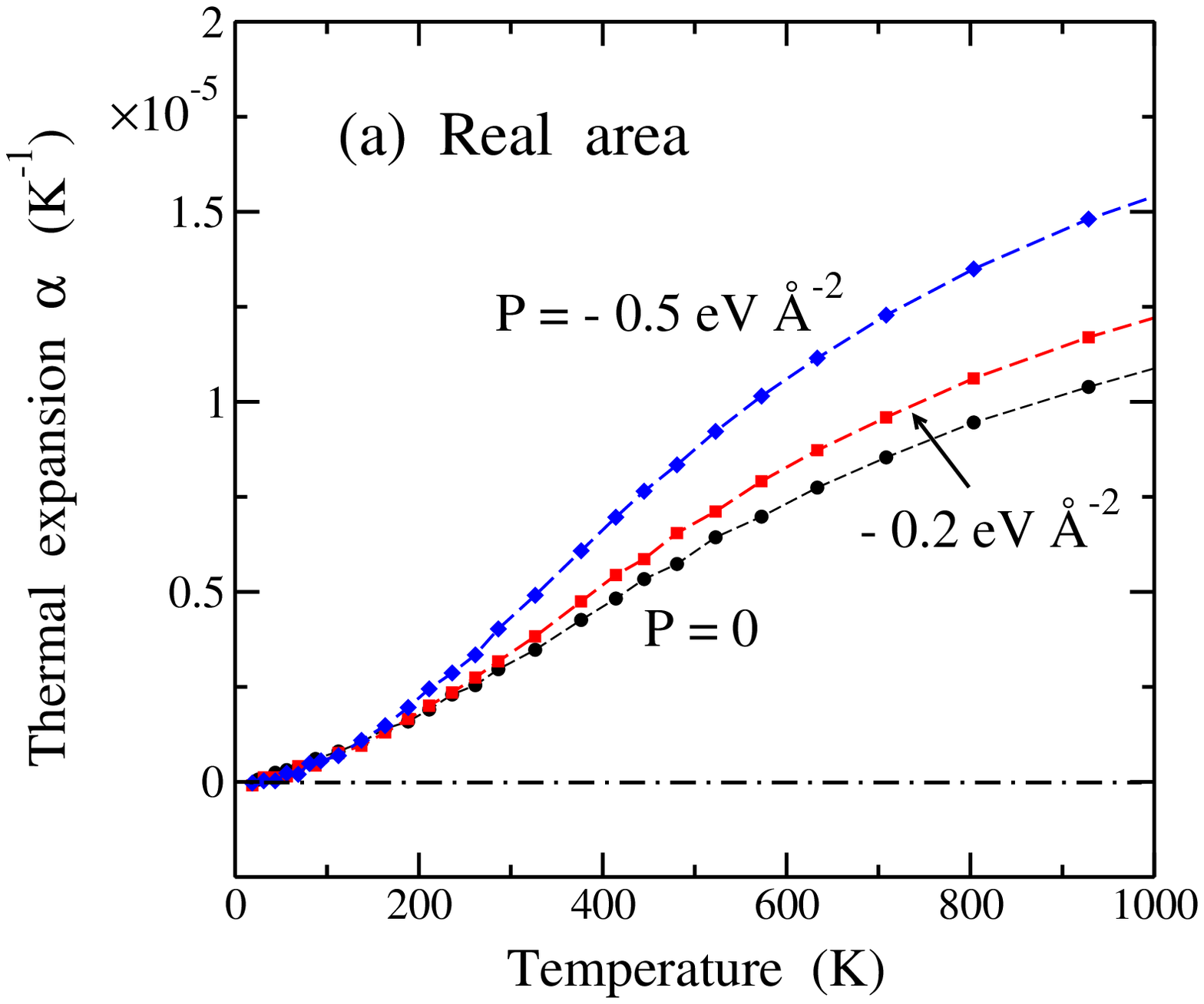}
~\vspace{-2.0cm}
\includegraphics[width= 8.0cm]{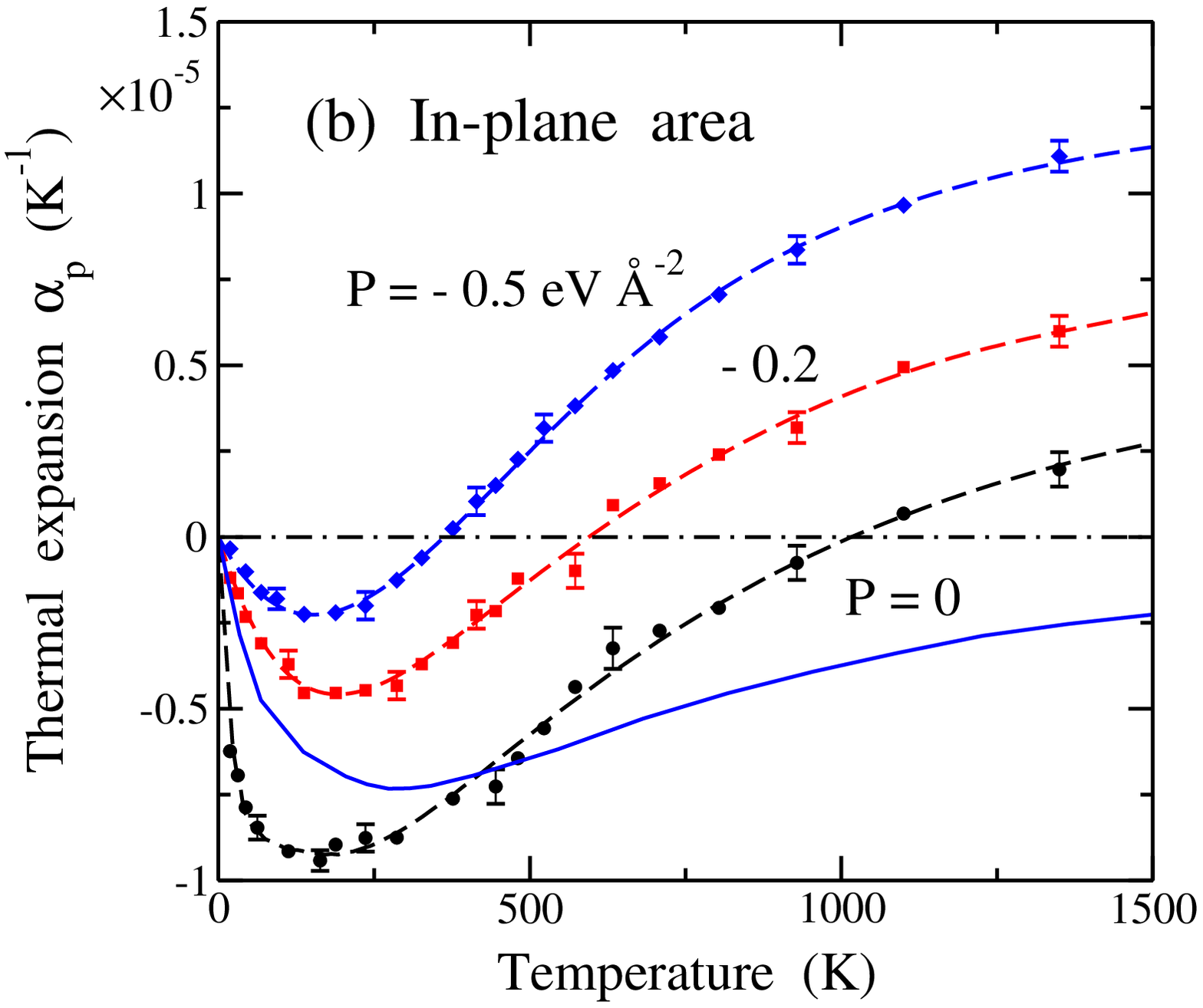}
\vspace{1.5cm}
\caption{(a) Thermal expansion coefficient $\alpha$ of graphene vs
temperature, as derived from the results of PIMD simulations.
(b) In-plane thermal expansion coefficient $\alpha_p$
of graphene vs temperature, obtained from numerical derivatives
of the area $A_p$.
In both panels, symbols represent data points for different stresses:
$P = 0$ (circles), $-0.2$ (squares), and $-0.5$ eV~\AA$^{-2}$ (diamonds).
Dashed lines are polynomial fits to the data points.
The solid line indicates the result obtained in Ref.~\onlinecite{mo05}
from density-functional perturbation theory.
}
\label{f6}
\end{figure}

To analyze the thermal expansion of graphene, and following our
definitions of the areas $A$ and $A_p$,
we will consider two different thermal expansion coefficients.
The first of them, $\alpha$, refers to changes in the real area:
\begin{equation}
 \alpha = \frac{1}{A}  \left( \frac{\partial A}{\partial T} \right)_P
\end{equation}
This coefficient takes mainly into account changes in the interatomic
distances, and is rather insensitive to bending of the graphene layer. 
The second coefficient, $\alpha_p$, is a measure of variations in
the in-plane area and has been widely used in the literature to 
analyze results of classical simulations and analytical calculations:
\begin{equation}
 \alpha_p = \frac{1}{A_p}
        \left( \frac{\partial A_p}{\partial T} \right)_P   \, .
\end{equation}

In Fig.~6 we present both thermal expansion coefficients,
as derived from our PIMD simulations for $P = 0$ (circles),
$-0.2$ (squares), and $-0.5$ eV \AA$^{-2}$ (diamonds).
Symbols are data points obtained from numerical derivatives of
the areas $A$ (panel a) and $A_p$ (panel b).
For these derivatives we took temperature intervals ranging from 10 K 
at low temperature to about 50 K at temperatures $T \sim 1000$~K.
In general, the statistical uncertainty (error bars) 
in the values of $\alpha_p$ obtained from numerical derivatives is 
larger than that found for $\alpha$, as a consequence of 
the larger fluctuations in $A_p$. The behavior of $\alpha$ shown in 
Fig.~6(a) is similar to that observed for most
3D materials, i.e., it goes to zero in the low-temperature limit
and increases for rising temperature so that $\alpha > 0$ for
$T > 0$.\cite{as76,he00c}
This is related to the fact that the in-plane vibrational modes
have positive Gr\"uneisen parameters when they are calculated with
respect to the real area $A$.
One observes in Fig.~6(a) that $\alpha$ is higher for larger tensile
stress. In fact, at 800 K the $\alpha$ value obtained for 
$P = -0.5$ eV \AA$^{-2}$ is 40\% larger than that corresponding to $P = 0$.
Viewing the thermal expansion as an anharmonic effect, 
this can be interpreted as an increase in anharmonicity for 
larger tensile stress. This is associated with an increase in area
$A$ and the corresponding decrease in the frequency of 
vibrational in-plane modes, which in turn causes a larger vibrational
amplitude and consequently a larger anharmonicity.

The in-plane thermal expansion coefficient $\alpha_p$ also converges
to zero in the low-temperature limit, but contrary to $\alpha$ 
it decreases for increasing temperature until reaching a minimum,
as shown in Fig~6(b). 
The negative value of $\alpha_p$ in the minimum approaches zero
for rising tensile stress. In fact, it goes from 
$-9.2 \times 10^{-6}$ K$^{-1}$ for $P = 0$ to
$-2.3 \times 10^{-6}$ K$^{-1}$ for $P = -0.5$ eV \AA$^{-2}$.
At higher $T$, $\alpha_p$ approaches zero and eventually becomes positive
at a temperature $T_0$, which changes appreciably with the applied stress. 
As a result, we find $T_0 = 1020$, 590 K, and 360($\pm 10$) K, 
for the three values of the stress $P$ presented in Fig.~6.

It is interesting to note that the difference $\alpha - \alpha_p$,
which vanishes at $T = 0$, rapidly increases for rising temperature,
and for $P = 0$ it takes a value $\approx 1.0 \times 10^{-5}$ K$^{-1}$ at
temperatures higher than 1000 K.
For larger tensile stress, this difference is smaller, since the
amplitude of out-of-plane vibrations is reduced, and $A_p$ is closer
to $A$. Thus, for $P = -0.2$ and $-0.5$ eV~\AA$^{-2}$, we find
at high temperatures $\alpha - \alpha_p$
$\approx 8 \times 10^{-6}$ K$^{-1}$ and
$6.5 \times 10^{-6}$ K$^{-1}$, respectively. 

The results presented here for $\alpha_p$ at zero external stress 
display a temperature dependence similar to 
those found earlier using other theoretical and 
experimental techniques.\cite{ji09,si14,ba09,yo11}
The solid line in Fig.~6(b) represents the thermal expansion
coefficient $\alpha_p$ obtained by Mounet and Marzari\cite{mo05}
from DFT combined with a QHA for the vibrational modes. Similar curves
$\alpha_p(T)$ were obtained from first-principles calculations 
in Refs.~\onlinecite{se14,ma17}, converging to negative values at
high temperature. This seems to be a drawback of this kind of 
calculations, which are optimal to obtain the total energy and 
electronic structure of the system, but the employed QHA
may be not precise enough to capture the important
coupling between in-plane and out-of-plane vibrational modes at
relatively high temperatures. This mode coupling controls the
in-plane thermal expansion.

Jiang {\em et al.}\cite{ji09} studied free-standing graphene using
a nonequilibrium Green's function approach, and obtained a minimum for 
$\alpha_p$ of about $-10^{-5}$ K$^{-1}$, similar to
our data for $P = 0$ displayed in  Fig.~6(b).  Moreover,
these authors found a crossover from negative to positive $\alpha_p$ 
for a temperature $T_0 \sim 600$ K, lower than the results of our PIMD
simulations.
An even lower temperature $T_0 \sim 400$ K has been found by using 
other theoretical techniques.\cite{si14,mi15b}
Room-temperature Raman measurements\cite{yo11} yielded
$\alpha_p = -8 \times  10^{-6}$ K$^{-1}$, whereas a value of
$-7 \times  10^{-6}$ K$^{-1}$ was derived from scanning electron 
microscopy.\cite{ba09}
We note, however, that the agreement between measurements with different 
techniques is not so good for the temperature dependence of $\alpha_p$, 
since the change in $\alpha_p$ at $T >$ 300~K is faster in the former 
case \cite{yo11} than in the latter.\cite{ba09}

The temperature dependence of $\alpha_p$ can be qualitatively
understood as a competition between two opposing factors.
On one side, the real area $A$ increases as $T$ is raised in the whole
temperature range considered here.
On the other side, bending of the graphene surface causes a decrease 
in its projection, $A_p$, onto the $xy$ plane.
For stress-free graphene at
temperatures $T \lesssim$ 1000~K, the decrease due to out-of-plane 
vibrations is larger than the thermal expansion of the real surface, 
and $\alpha_p < 0$.
For $T \gtrsim$ 1000~K, the thermal expansion of $A$ dominates over
the contraction of $A_p$ associated to out-of-plane atomic motion,
and thus $\alpha_p > 0$.
For graphene under tensile stress, the amplitude of out-of-plane
vibrations is smaller, so the decrease in $A_p$ is also smaller and the
region of negative $\alpha_p$ is reduced.

Our results for the thermal expansion of graphene, derived from 
atomistic simulations, can be related with an analytical formulation of 
crystalline membranes in the continuum limit, for which the
relation between $A$ and $A_p$ may be written as\cite{wa09,ra17}
\begin{equation}
  A = \int_{A_p} dx \, dy \, \sqrt{1 + (\nabla h(x,y))^2}  \; .
\label{aap}
\end{equation}
Here $h(x,y)$ is the height of the membrane surface, i.e. the distance to 
the reference $xy$ plane.
In this classical approach, the difference $A - A_p$ 
may be calculated by Fourier transformation 
of the r.h.s. of Eq.~(\ref{aap}).\cite{sa94,ch15,ra17}
In this procedure one needs to consider a dispersion relation 
$\omega_{\rm ZA}({\bf k})$ for out-of-plane modes (ZA flexural band), 
where ${\bf k} = (k_x, k_y)$ are 2D wavevectors. 
The frequency dispersion in this acoustic band is well approximated 
by the expression $\rho \, \omega_{\rm ZA}^2 = \sigma k^2 + \kappa k^4$,
consistent with an atomic description of graphene\cite{ra16}
($k = |{\bf k}|$; $\rho$, surface mass density; 
$\sigma$, effective stress; $\kappa$, bending modulus).
Then, for effective stress $\sigma > 0$, which is the case at finite 
temperatures, even for zero external stress ($P = 0$),
one finds\cite{ra16,ra17}
\begin{equation}
 A = A_p \left[ 1 + \frac {k_{B} T} {8 \pi \kappa}
      \ln \left(1 + \frac{2 \pi \kappa} {\sigma A_p} \right) \right]  \:.
\label{aap3}
\end{equation}
This expression was obtained in the classical limit, so it does not 
take into account atomic quantum delocalization. Nevertheless, it is
expected to be a good approximation to our quantum calculations
at relatively high temperature, $T \gtrsim \Theta_D$, with  
$\Theta_D \sim$ 1000 K the Debye temperature corresponding to 
out-of-plane vibrations in graphene.\cite{te09,po11}

We note that both $\kappa$ and $\sigma$ change with temperature and
stress, so that it is not straightforward to write down an analytical 
formula for $\alpha - \alpha_p$ from a temperature derivative 
of Eq.~(\ref{aap3}).
For given temperature $T$ and stress $\sigma = \sigma_0 - P$
we can write $A - A_p = \delta A_p T$, $\delta$ being a parameter
derived from Eq.~(\ref{aap3}), which is a good approximation for
the difference $\alpha - \alpha_p$.
The effective stress $\sigma_0$ appears at finite temperatures for zero
external stress ($P = 0$), and vanishes for $T \to 0$ in the classical
limit.\cite{ra16}
Introducing the finite-temperature values of $\sigma_0$ and $\kappa$ given 
earlier,\cite{ra16,ra17} we find $\delta$ = $1.0 \times 10^{-5}$ K$^{-1}$,
$6.6 \times 10^{-6}$ K$^{-1}$, and $5.1 \times 10^{-6}$ K$^{-1}$
for $P = 0$, $-0.2$, and $-0.5$ eV \AA$^{-2}$, respectively.
These values are close to those found for $\alpha - \alpha_p$ from
our PIMD simulations (see above). 
However, the difference between both sets of results increases for 
rising tensile stress. Thus, for $P = -0.5$ eV \AA$^{-2}$ 
our simulations yield $\alpha - \alpha_p = 6.5 \times 10^{-6}$~K$^{-1}$, 
larger than the prediction based on Eq.~(\ref{aap3}), which is based on
harmonic vibrations. We observe that the importance of anharmonic effects 
(including also quantum corrections) is manifested
more clearly for large tensile stresses.

\begin{figure}
\vspace{-0.7cm}
\includegraphics[width=8.0cm]{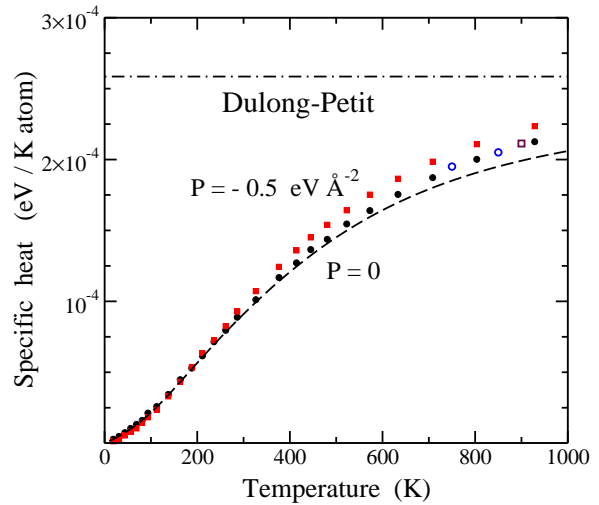}
\vspace{-0.5cm}
\caption{Specific heat of graphene as a function of temperature,
as derived from PIMD simulations for $P = 0$ (circles) and
$-0.5$ eV \AA$^{-2}$ (squares).
Error bars of the data points are less than the symbol size.
The dashed line is the specific heat $c_v$ obtained from the six phonon
bands corresponding to the LCBOPII potential in a harmonic approximation
($P = 0$).
Open symbols indicate results for $c_v$ derived from DFT-type calculations
at $T$ = 900 K (square, Ref.~\onlinecite{ma16}),
850 K and 750 K (circles, Ref.~\onlinecite{ma12}).
The horizontal dashed-dotted line represents the classical
Dulong-Petit limit.
}
\label{f7}
\end{figure}

\section{Specific heat}

We have calculated the specific heat of graphene monolayers as 
a temperature derivative of the enthalpy $H$ derived from our
PIMD simulations (see Sec.~III.D): $c_p(T) = dH(T) / dT$.
One may ask if our simulations can yield reliable results for 
this thermodynamic variable, mainly because electronic contributions 
to $c_p$ are not taken into account in our numerical procedure.
This is, however, not a problem for the actual precision reached
in our calculations, as the electronic part is much less than 
the phonon contribution, actually considered in our method. 
The former was estimated in various works, and turns out to be 
three or four orders of magnitude less than the phonon part 
in the temperature range considered here.\cite{be96,ni03,fo13}

In Fig.~7 we display the specific heat of graphene as a function 
of temperature for $P = 0$ (circles) and $-0.5$ eV \AA$^{-2}$ (squares),
as obtained from a numerical derivative of the enthalpy. 
For comparison, the classical Dulong-Petit specific heat is shown as
a dashed-dotted line ($c_v^{\rm cl} = 3 k_B$).
At room temperature the specific heat of graphene is about three times
smaller than the classical limit, and $c_p$ is still appreciably 
lower than this limit at $T = 1000$~K. This is in line with
the magnitude of the Debye temperature of graphene, which amounts
to about 1000~K for out-of-plane modes\cite{te09,po11} and
$\Theta_D \gtrsim 2000$~K for in-plane vibrations.\cite{te09,po12b}
In Fig.~7 we also present results for $c_v$ at $P = 0$ taken from earlier 
DFT calculations combined with a QHA: 
$T$ = 900 K from Ref.~\onlinecite{ma16} (open square), 
850 K and 750 K from Ref.~\onlinecite{ma12} (open circles).
These data agree well with the results of our PIMD simulations.

For a comparison with our numerical results for $c_p(T)$
at zero and negative stress,
we present a harmonic approximation (HA) for the lattice vibrations.
This approximation is expected to be rather accurate at low temperatures,
provided that a reliable description for the phonon frequencies 
is used. A comparison between results of the HA and PIMD simulations
yields an estimate of anharmonic effects in the specific
heat of graphene, given that both kinds of calculations employ
the same interatomic potential.

The HA assumes constant frequencies for the graphene vibrational modes 
(calculated for the minimum-energy configuration), and does not take 
into account changes in the areas $A$ and $A_p$ with temperature. 
Thus, it will give us the constant-area specific heat per atom, 
$c_v(T) = d E(T) / d T$, which for a cell with $N$ atoms is given by
\begin{equation}
 c_v(T) = \frac {k_B}{N} \sum_{r,\bf k}
   \frac { \left[ \frac12 \beta \hbar \, \omega_r({\bf k}) \right]^2 }
   { \sinh^2 \left[ \frac12 \beta \hbar \, \omega_r({\bf k}) \right] } \, ,
\label{cvn}
\end{equation}
where the index $r$ ($r$ = 1, ..., 6) indicates the six phonon bands
of graphene (ZA, ZO, LA, TA, LO, and TO),\cite{mo05,ka11,wi04}
and the sum in ${\bf k}$ runs over wavevectors
${\bf k} = (k_x, k_y)$ in the hexagonal Brillouin zone,
with discrete ${\bf k}$ points spaced by $\Delta k_x = 2 \pi / L_x$ and
$\Delta k_y = 2 \pi / L_y$.\cite{ra16}
For increasing system size $N$, one has new long-wavelength modes 
with an effective cut-off
$\lambda_{max} \approx L$, with $L = (N A_p)^{1/2}$,
and the minimum wavevector is
$k_0 = 2 \pi / \lambda_{max}$ (i.e., $k_0 \sim N^{-1/2}$).

The dashed line in Fig.~7 was calculated with the HA using
Eq.~(\ref{cvn}), with the frequencies $\omega_r({\bf k})$ ($r$ = 1, ..., 6)
obtained from diagonalization of the dynamical matrix corresponding
to the LCBOPII potential.
Results of the simulations for $P = 0$ follow closely the HA up to
about 400~K, and they become progressively higher than the dashed line
at higher temperatures, for which anharmonic effects become more important.
At room temperature ($T = 300$~K) we obtained 
$c_v = 9.1 \times 10^{-5}$ eV/K atom from the HA vs
a value of $9.3(\pm 0.1) \times 10^{-5}$ eV/K atom derived from 
PIMD simulations.  Both values are a little higher than that found 
by Ma {\em et al.}\cite{ma12} from DFT calculations 
($c_v = 8.6$ J/K mol, i.e. $8.9 \times 10^{-5}$ eV/K atom).
The difference between results obtained for $P = 0$ 
from {\em ab-initio} calculations combined with a QHA and those 
presented here for a simpler HA are due to two main reasons. 
The first reason is that in the HA one does not take into account 
changes of frequencies with the temperature, and the second is the
relative inaccuracy of the phonon bands derived from the considered
effective potential far from the $\Gamma$ point, as compared with
those obtained from DFT calculations. The HA provides, however, a
consistency check for the results of the PIMD simulations at low
temperature, as discussed below.

As shown in Sec.~III, a part of the internal energy (and the enthalpy)
corresponds to the elastic energy $E_{\rm el}$,
i.e., to the cost of increasing the area $A$ of graphene by thermal
expansion or an applied tensile stress.
The contribution of this energy to the specific heat is given
by $d E_{\rm el} / d T$, which is not included in the HA.
This contribution can be calculated from the results of our PIMD
simulations displayed in Fig.~3.
For $P = 0$, it amounts to 2.3 and $5.5 \times 10^{-6}$ eV/K atom at
500 and 1000 K, which means a nonnegligible increase in the specific 
heat with respect to the pure HA. This increase is especially visible at
$T \gtrsim 500$~K, capturing part of the anharmonicity of the system.
In fact, at $T = 1000$~K it accounts for a 45\% of the difference
between the results derived from PIMD simulations and
the HA presented in Fig.~7. The rest of that
difference is associated to anharmonicity of the lattice 
vibrations.

\begin{figure}
\vspace{-0.7cm}
\includegraphics[width=8.0cm]{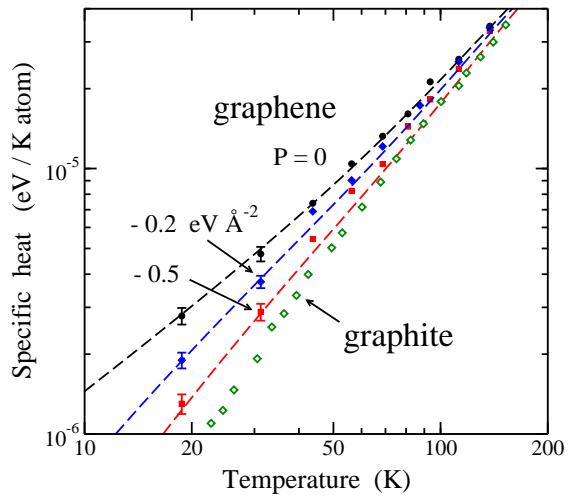}
\vspace{-0.5cm}
\caption{Specific heat of graphene as a function of temperature.
Symbols represent results derived from PIMD simulations for
$P = 0$ (circles), $-0.2$ (diamonds), and $-0.5$ eV \AA$^{-2}$ (squares).
Lines were derived from a HA based on the six phonon bands corresponding
to the LCBOPII potential for N = 61440 (see text for details).
Experimental data for graphite obtained by
Desorbo and Tyler\cite{de53} are shown as open diamonds.
}
\label{f8}
\end{figure}

In our present context, the most relevant effect of an applied stress $P$
in the phonon bands is a change in the low-frequency region of the ZA modes,
for which
\begin{equation}
  \omega_{\rm ZA}({\bf k})^2 = \omega^0_{\rm ZA}({\bf k})^2 -
          \frac{P}{\rho} \, k^2
\label{omega2}
\end{equation}
The zero-stress band $\omega^0_{\rm ZA}({\bf k})$ calculated for the
minimum-energy structure (area $A_0$), follows for small $k$:
$\rho \, \omega^0_{\rm ZA}({\bf k})^2 \approx \kappa k^4$.
Thus, for $P < 0$ the small-$k$ region is dominated by the quadratic term
(linear in $P$) in Eq.~(\ref{omega2}), and
$\omega_{\rm ZA}({\bf k}) \approx \sqrt{-P/\rho} \; k$
for $k \ll$ 1~\AA$^{-1}$.
The specific heat of stressed graphene in the HA has been calculated
by taking the frequencies $\omega_{\rm ZA}({\bf k})$ derived from 
Eq.~(\ref{omega2}). 

To obtain deeper insight into the low-temperature dependence of the
specific heat, $c_p(T)$ is shown in Fig.~8 in a logarithmic plot.
Symbols indicate results  derived from PIMD simulations for 
$P = 0$ (circles), $-0.2$ (squares), and $-0.5$ eV \AA$^{-2}$ (diamonds).
With our present method, we cannot give reliable values of $c_p$ for 
temperatures lower than 15~K, mainly due to the very large computation times
required for Trotter numbers $N_{\rm Tr} > 500$.
Dashed lines show $c_v(T)$ derived from the harmonic approximation 
using Eq.~(\ref{cvn}) for a large cell size ($N = 61440$ atoms). Such
a large $N$ cannot be reached in our quantum simulations, and gives us
a useful reference where finite-size effects are minimized.
The lines corresponding to the HA follow closely the data points derived 
from the simulations at low temperatures. For $P = -0.5$ eV \AA$^{-2}$ 
one observes that the PIMD results become progressively higher than 
the HA line at $T > 50$~K. Such a difference is 
almost unobservable for $P = -0.2$ eV \AA$^{-2}$, and disappears for
$P = 0$. It is clear that the HA describes well the specific heat in
the stress-free case at $T < 200$~K, and becomes less accurate as 
tensile stress and/or temperature increase.

The low-temperature behavior of the heat capacity can be further 
analyzed by considering a continuous model for frequencies and
wavevectors, as in the well-known Debye model for solids.\cite{as76}
At low-temperatures, $c_v$ is mainly controlled by the contribution of
acoustic modes with small $k$. For graphene, these are TA and LA modes
with $\omega_r \propto k$, and ZA flexural modes with quadratic 
dispersion ($\omega_r \propto k^2$) for negligible $\sigma$ at
$P = 0$ and low $T$, whereas it is linear in $k$ for $P < 0$.
In general, the low-$T$ contribution of a phonon branch with dispersion
relation $\omega_r \propto k^n$ can be approximated by replacing
the sum in Eq.~(\ref{cvn}) by an integral, which yields
$c_v^r \sim T^{2/n}$ (see Ref.~\onlinecite{he18}).
We note that for $d$-dimensional systems one finds
an exponent $d/n$.\cite{ho01,po02}

Summarizing the above comments, the low-temperature contributions 
of the relevant phonon branches to the specific heat
(those with $\omega(k) \to 0$ for $k \to 0$), one has for graphene 
$c_v \sim T^{\mu}$ close to $T = 0$, with $\mu = 1$ for 
$P = 0$ and $\mu = 2$ in the presence of a tensile stress ($P < 0$).
This is in fact what we find for $c_v$ from the HA at low temperature
in our calculations for a large graphene supercell.  
To see the convergence of $\mu$ to its low-$T$ value,
we have calculated this exponent as a function of temperature 
from a logarithmic derivative: $\mu = d \ln c_v / d \ln T$. 
Close to $T = 0$ we obtain the values
expected from our discussion above, but in the region displayed in
Fig.~8, $\mu$ has not yet reached the zero-temperature value.
In fact, for $T = 10$~K we find for $\mu$ the values 
1.05, 1.59, and 1.85, for $P = 0$, $-0.2$, and $-0.5$ eV \AA$^{-2}$,
respectively. The first of them is already close to its 
low-temperature limit ($\mu = 1$), but the exponents for
graphene under stress are still somewhat lower than their zero-$T$
limit ($\mu = 2$). We note that the larger the tensile stress 
(larger $\sigma = \sigma_0 - P$), the closer is $\mu$ to 2 at 
a relatively low $T$, since the wavenumber region 
where the linear term dominates in the $\omega(k)$ 
dispersion of the ZA band is larger
($\rho \omega_{\rm ZA}^2 \approx \sigma k^2$).
For the results of our PIMD simulations we note that,
although one can obtain an exponent $\mu$ from a linear fit in 
the logarithmic plot of Fig.~8 (in particular the fit is rather good 
for $T < 100$ K), it is true that the actual slope is slowly changing 
in this temperature range.

Alofi and Srivastava\cite{al14} calculated the specific heat of
few-layer graphene by using a semicontinuum model and analytical
expressions for phonon dispersion relations. In particular, for 
single-layer graphene they found a temperature dependence
$c_v \sim T^{1.1}$ for $T \gtrsim 10$~K, with an exponent that 
coincides with the mean $\mu$ value derived from our results 
for $P = 0$ in the region from 10 to 50 K.

For comparison with our results for graphene, we also present 
in Fig.~8 experimental data for the specific heat $c_p$ of
graphite, obtained by Desorbo and Tyler.\cite{de53}
For graphite, the dependence of $c_p$ on temperature has been 
analyzed in detail over the years.\cite{ko51,kr53,kl53,ni72}
For $T < 10$~K, $c_p$ rises as $T^3$ (a temperature region not shown 
in Fig.~8), as in the Debye model, and for temperatures 
between 10 and 100~K, it increases as $T^2$ ($\mu = 2$).
The major difference with stress-free graphene is that 
in graphite the dominant 
contribution to $c_p$ in this temperature range arises from phonons 
with a linear dispersion relation for small $k$ ($\omega \sim k$). 
At room temperature the experimental specific heat of graphite is 
$8.90 \times 10^{-5}$ eV/K atom (8.59 J/K mol), somewhat less 
than the result for graphene derived from our PIMD simulations for
$P = 0$: $c_p = 9.3(\pm 0.1) \times 10^{-5}$ eV/K atom.

To compare with the results for $c_p$ of graphene derived from our PIMD
simulations, we have also calculated the specific heat $c_v$ from 
constant-$A_p$ simulations.  For each considered temperature, 
we took the equilibrium area $A_p$ obtained in 
the isothermal-isobaric simulations, 
and calculated $c_v$ as a numerical derivative of the internal energy 
\begin{equation}
   c_v = \left( \frac{ \Delta E } { \Delta T }  \right)_{A_p} 
\end{equation}
from temperature increments $\Delta T$ (both positive and negative).
One expects that $c_v \leq c_p$ at any temperature, but the difference
between them turns out to be less than the statistical error bars
of our results. A realistic estimation of this difference can be
obtained from the thermodynamic relation\cite{la80,he18}
\begin{equation}
  c_p - c_v = \frac{T \alpha_p^2 A_p} {\chi_p}  \,
\label{cpcv}
\end{equation}
where $\chi_p$ is the in-plane isothermal compressibility (see Sec.~VI).
Using Eq.~(\ref{cpcv}), we find that $c_p - c_v$ is at least two orders 
of magnitude less than the $c_p$ values given above for the tensile 
stresses and temperatures considered here.

We finally note that a calculation of the low-$T$ specific
heat of solids using this kind of path-integral simulations is 
in general not straightforward.
The verification of the Debye law $c_p \sim T^3$ has been a challenge 
for path-integral simulations of 3D materials, because of the effective
low-frequency cut-off corresponding to finite simulation 
cells.\cite{no96,ra06}
This kind of calculations at relatively low temperatures are
in principle more reliable for 2D materials due to two different reasons.
The first reason is that the length of the cell sides scales as 
$N^{1/d}$, and the minimum wavevector $k_0$ available
in the simulations is $k_0 \sim N^{-1/d}$. 
For a simulation cell including $N$ atoms, $k_0$ is less for 2D than 
for 3D materials, so that the low-frequency region is represented better 
for $d = 2$, and therefore the low-temperature regime will be also
better described. 
The second reason is that the internal energy (or enthalpy) for 
graphene rises as $T^{\mu + 1}$ with $\mu = 2$ or 3. 
At low $T$, this increase is faster than the usual phonon contribution 
to the enthalpy in 3D materials ($H \sim T^4$), and consequently it is
more readily observable (less relative statistical noise) 
for 2D materials such as graphene.

\begin{figure}
\vspace{-0.7cm}
\includegraphics[width=8.0cm]{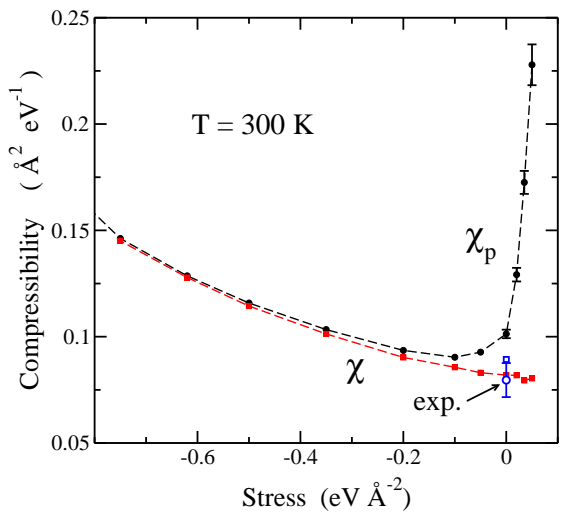}
\vspace{-0.5cm}
\caption{Stress dependence of the isothermal compressibilities
$\chi$ (squares) and $\chi_p$ (circles) of graphene at $T = 300$ K,
as derived from PIMD simulations.
Error bars, when not shown, are in the order of the symbol size.
Open symbols indicate results derived from AFM indentation experiments:
a square from Ref.~\onlinecite{lo15b} and a circle from
Ref.~\onlinecite{le08} (the error bar corresponds to one standard
deviation for several measurements).
}
\label{f9}
\end{figure}

\section{Compressibility}

The in-plane isothermal compressibility is defined as
\begin{equation}
   \chi_p = - \frac{1}{A_p} 
            \left( \frac{\partial A_p}{\partial P} \right)_T   \, .
\label{chip1}
\end{equation}
The variables in the r.h.s. of this equation refer to 
in-plane quantities, as the pressure $P$ in our isothermal-isobaric
ensemble is a variable conjugate to the in-plane area $A_p$.
$\chi_p$ has been calculated here by using the fluctuation
formula\cite{la80,ra17}
\begin{equation}
   \chi_p = \frac{N \sigma_p^2}{k_B T A_p}
\label{chip2}
\end{equation}
where $\sigma_p^2$ are the mean-square fluctuations of the
area $A_p$ obtained in the simulations.
This expression is more convenient for our purposes than obtaining
$(\partial A_p / \partial P)_T$, since a calculation of this derivative
by numerical methods requires additional simulations at nonzero stresses.
We have verified at some selected temperatures and pressures that
both procedures give the same results for $\chi_p$
(taking into account the statistical error bars).
Similarly, for the real area $A$ one can define a compressibility
$\chi = - (\partial A / \partial P)_T / A$, which will be related to
the fluctuations of the real area $A$.

In Fig.~9 we present the stress dependence of the compressibilities 
$\chi_p$ (circles) and $\chi$ (squares) of graphene, as derived from our 
PIMD simulations at 300 K.
At $P = 0$, one finds $\chi_p > \chi$, as a consequence of the larger
fluctuations in the in-plane area $A_p$.
In this figure we have included some points corresponding to (small) 
compressive stresses $P > 0$, to remark the very different behavior 
of $\chi_p$ and $\chi$ in this region. 
$\chi$ follows a regular dependence, in the sense that it displays 
a smooth change as $P$ is varied in the considered stress region.
The in-plane compressibility $\chi_p$, however, increases fast for
$P > 0$, which is consistent with an eventual divergence at a
critical stress $P_c$.
This divergence of $\chi_p$ shows the same fact as
the vanishing of the in-plane bulk modulus $B_p$
(the inverse of $\chi_p$) discussed in Ref.~\onlinecite{ra17} from
classical calculations. 
Close to this critical compressive stress the planar morphology 
of graphene becomes unstable due to
large fluctuations in the in-plane area, which is related to
the onset of imaginary frequencies in the ZA phonon bands for $P > P_c$.
A detailed characterization of this wrinkling transition at $P_c$ is
out of the scope of this paper and will be investigated further
in the near future. It is not yet clear whether quantum effects may
be relevant or not for a precise definition of $P_c$.
It is remarkable that the compressibility $\chi$ derived from the real
area does not display any abrupt change in the region close 
to the critical stress, 
as both the area $A$ and its fluctuations are rather insensitive to
the corrugation of the graphene surface imposed by compressive stresses.
For large tensile stress, $\chi$ and $\chi_p$ converge one to the
other, as shown in Fig.~9, since the amplitude of out-of-plane 
vibrations becomes smaller and the real graphene surface is closer to
the reference $xy$ plane.

For comparison with the results of our simulations, we also present in
Fig.~9 data for the compressibility of graphene at $P = 0$ derived from 
atomic force microscopy (AFM) indentation experiments. 
We have transformed the values for the Young
modulus $Y$ given in Refs.~\onlinecite{le08,lo15b} to compressibility
by using the expression $\chi = 2 (1 - \nu) / Y$, with a Poisson
ratio $\nu = 0.15$.\cite{ra17} The resulting compressibilities are plotted
as open symbols: a square\cite{lo15b} and a circle\cite{le08}
(the error bar indicates one standard deviation for several measurements). 
We note that interferometric profilometry experiments\cite{ni15} 
have revealed that close to $P = 0$ much smaller values of the Young
modulus (much larger compressibilities) can be found for graphene.
These authors have suggested that this apparent discrepancy can be 
associated to the difference between compressibilities of the real and
projected areas, as discussed here. It seems that some experimental
techniques can measure one of them, while other techniques may be sensitive
to the other. This is an ongoing discussion that should be elucidated 
in the near future.\cite{ra17,ni17}

A thermodynamic parameter related to the thermal expansion $\alpha_p$ and 
compressibility $\chi_p$ is the dimensionless Gr\"uneisen parameter 
$\gamma$, defined as
\begin{equation}
  \gamma = \frac{ \sum_{r {\bf k}} \gamma_{r {\bf k}} \, c_{v r}({\bf k}) }
            { \sum_{r {\bf k}} c_{v r}({\bf k}) }  \, ,
\label{gamma1}
\end{equation}
where $c_{v r}({\bf k})$ is the contribution of mode $(r {\bf k})$ to
the specific heat, and $\gamma_{r {\bf k}}$ are mode-dependent
Gr\"uneisen parameters.\cite{as76,mo05}
The overall $\gamma$ can be related with our in-plane variables 
in graphene by using the thermodynamic relation\cite{as76,he18}
\begin{equation}
       \gamma = \frac{\alpha_p A_p}{\chi_p c_v}  \, .
\label{gamma2}
\end{equation}
For $P = 0$, the results of our PIMD simulations yield
$\gamma = -2.2$ at 300~K, and at 1000~K, $\gamma \approx 0$
within the precision of the numerical results.
At $T = 300$~K, $\alpha_p$ and $\gamma$ are negative because
the mode-dependent Gr\"unesien parameter of the out-of-plane 
ZA modes is negative.\cite{ba11,mo05,ma16}
However, for $T > 1000$~K, this negative contribution is 
dominated by the positive sign of $\gamma_{r {\bf k}}$ for 
in-plane modes, which are excited at these temperatures, so that
$\alpha_p$ and $\gamma$ are positive.

At room temperature ($T = 300$~K) we find $\gamma = -1.2$ and $-0.24$,
for $P = -0.2$ and $-0.5$ eV~\AA$^{-2}$, respectively. 
Although these values of $\gamma$ are still negative, they are
very different from the zero-stress result ($\gamma = -2.2$).
Looking at Eq.~(\ref{gamma2}), the main reason for this important
change in the Gr\"unesien parameter with tensile stress is the
large variation in the in-plane thermal expansion coefficient $\alpha_p$,
which takes values of $-8.5$, $-3.9$, and $-1.0 \times 10^{-6}$~K$^{-1}$
for $P = 0$, $-0.2$, and $-0.5$ eV~\AA$^{-2}$, respectively (see Fig.~6(b)).
From the point of view of the phonon contributions to $\gamma$, 
the small negative value found for $P = -0.5$ eV~\AA$^{-2}$
indicates that the relative contribution of phonons with negative
$\gamma_{r {\bf k}}$ (out-of-plane ZA modes) is at room temperature
less important than for $P = 0$. This is due to an increase in 
frequency of small-$k$ modes 
($\omega_{\rm ZA} \sim \sqrt{- P / \rho} \; k$),
which causes a reduction in its corresponding $c_{v r}({\bf k})$
at 300~K (see Eq. ~(\ref{gamma1})).

\section{Summary}

In this paper we have presented and discussed results of PIMD simulations 
of graphene in a wide range of temperatures and tensile stresses.
This technique has allowed us to quantify several structural and 
thermodynamic properties, with particular emphasis on the thermal
expansion and specific heat.
Nuclear quantum effects are clearly appreciable in these variables,
even at $T$ higher than room temperature.
Zero-point expansion of the graphene layer due to nuclear quantum
motion is not negligible, and amounts to about 1\% of the area $A$. 
This zero-point effect decreases as tensile stress is increased.

The thermal contraction of graphene discussed in the literature turns out 
to be a reduction of the in-plane area $A_p$ ($\alpha_p < 0$), caused by 
out-of-plane vibrations, but not a decrease in the real area $A$.
In fact, the difference $A - A_p$ rises as temperature increases,
since the amplitude of those vibrations grows.
On one side, the in-plane thermal expansion $\alpha_p$ is negative at 
low temperature, and becomes positive for $T \gtrsim$ 1000~K.
On the other side, the thermal expansion $\alpha$ of the real area is 
positive for all temperatures and tensile stresses discussed here.
The thermal contraction of $A_p$ is smaller as the tensile stress 
increases, due to a reduction in the amplitude of out-of-plane vibrations.
Our PIMD simulations give $\alpha_p < 0$ at low $T$ for stresses
so high as $-0.5$ eV \AA$^{-2}$ ($-8$~N/m). 
For this stress, $\alpha_p$ becomes positive at $T \sim 350$~K.

The anharmonicity of the vibrational modes is clearly noticeable 
in the behavior of $A$ and $A_p$. The increase in real area $A$ for
rising $T$ ($\alpha > 0$) is a consequence of anharmonicity of
in-plane modes, similar to the thermal expansion in most 3D solids.
Moreover, the peculiar dependence of $\alpha_p$ (negative at low $T$
and positive at high $T$) is an indication of the coupling 
between in-plane and out-of-plane modes.

Other thermal properties of graphene can be well described
by an HA at relatively low $T$, once the frequencies of the vibrational 
modes are known for the classical equilibrium geometry at $T = 0$.
This is the case of the specific heat, for which our PIMD results
indicate that the effect of anharmonicity appears gradually
for temperatures $T \gtrsim$ 400~K.
Part of this anharmonicity is due to the elastic energy $E_{\rm el}$,
associated to the expansion of the actual graphene sheet, which
is not taken into account by the HA.
At low $T$ the specific heat can be described as $c_p \sim T^{\mu}$.
At the lowest temperatures studied here ($T \gtrsim$ 10~K) we find
an exponent $\mu$ = 1.05 and 1.85, for $P = 0$ and 
$-0.5$ eV \AA$^{-2}$, respectively.  These values are close to the
corresponding low-temperature ($T \to 0$) values, i.e., 
$\mu$ = 1  for stress-free and $\mu$ = 2 for stressed graphene.

We note the consistency of the simulation results with the 
principles of thermodynamics, in particular with the third law.
This means that thermal expansion coefficients have to vanish 
for $T \to 0$, as found for both the in-plane area $A_p$ and the real 
area $A$ derived from our simulations for $P = 0$ and $P < 0$.
The same happens for the specific heat, i.e., $c_p \to 0$ for
$T \to 0$.

\begin{acknowledgments}
The authors acknowledge the help of J. H. Los in the implementation 
of the LCBOPII potential.
This work was supported by Direcci\'on General de Investigaci\'on,
MINECO (Spain) through Grants FIS2012-31713 and FIS2015-64222-C2.
\end{acknowledgments}

\end{document}